\documentclass[12pt]{article}
\usepackage{geometry} 
\usepackage{graphicx} 
\usepackage{amsmath} 
\usepackage{amsfonts} %
\usepackage{cite}          
\usepackage{hyperref}  

\usepackage{tikz}
\usepackage{pgfplots}
\usetikzlibrary{decorations.markings}
\usetikzlibrary{patterns}
\def\Czzdash{
\begin{tikzpicture}[scale=0.5,>=latex,decoration={
markings,mark=between positions 0.05 and 0.99 step 0.33cm with {\arrow{to}}}]
 \filldraw (0,0) circle (0.5pt) node[above left=-5pt] {\fontsize{5pt}{0pt}\selectfont\mbox{$z'$}};
 \draw[postaction={decorate}] (0,0) circle (12pt);
 \node at (0.45,0.45) {\fontsize{5pt}{0pt}\selectfont\mbox{$ z$}};
 \end{tikzpicture}
}
\def\cc{\mbox{c}_{\scalebox{0.36}{\bf CFT}}}
\def\ccs{\mbox{\scriptsize c}_{\scalebox{0.2}{\bf CFT}}}
\def\ccb{\mbox{\={c}}_{\scalebox{0.36}{\bf CFT}}}
\def\ccbs{\mbox{\scriptsize \={c}}_{\scalebox{0.2}{\bf CFT}}}
\title{Time development of conformal field theories associated with  $L_{1}$ and $L_{-1}$ operators}
\author{Tsukasa TADA}
\date{
}
\begin{document}
\maketitle
\centerline{\it RIKEN Nishina Center for Accelerator-based Science \&} 
\centerline{\it Interdisciplinary Theoretical and Mathematical Sciences (iTHEMS)}
\centerline{\it Wako, Saitama 351-0198, Japan}
\renewcommand{\L}{{\cal L}}
\renewcommand{\l}{\mathfrak l}
\begin{abstract}
In this study, we examined consequences of unconventional time development of two-dimensional conformal field theory induced by the $L_{1}$ and $L_{-1}$ operators, employing the formalism previously developed in a study of sine-square deformation. We discovered that the retainment of the Virasoro algebra requires the presence of a cut-off near the fixed points. The introduction of a scale by the cut-off makes it possible to recapture the formula for entanglement entropy in a natural and straightforward manner.
\end{abstract}

\section{Introduction}
In \cite{Ishibashi:2015jba,Ishibashi:2016bey}, a formalism was conceived in which unconventional time developments other than the radial time development can be used to study conformal field theories (CFTs). In particular, the sine-square deformation (SSD) \cite{Gendiar:2008udd} of  two-dimensional (2D) CFT \cite{Katsura:2011ss} has been explained with this formalism as a particular time development called ``dipolar quantization''. There are earlier studies \cite{1104.1721,1108.2973} on SSD, studies in the context of string theory \cite{Tada:2014kza,Tada:2014wta}, and  more recent studies on this topic \cite{Okunishi:2015dfa,Okunishi:2016zat}. An elegant generalization has also been developed by Wen, Ryu and Ludwig \cite{Wen:2016inm}, which involves the entanglement Hamiltonian and other interesting deformations of 2D CFT. The most recent studies on the subject include Refs. \cite{Tamura:2017vbx,Tada:2017wul,Wen:2018vux,Wen:2018agb,Kishimoto:2018ekq,Wen:2018svb}.

In the present study, we examine  the case of a particular time development which was left for further study in Ref. \cite{Ishibashi:2016bey}. The time development in question can be achieved using
\begin{equation}
L_{1}+L_{-1}+{\bar L}_{1}+{\bar L}_{-1} ,  \label{eqn:L1L-1Hamiltonian}
\end{equation}
as the time-development operator, instead of $L_{0}+{\bar L}_{0}$. The holomorphic part of Eq. (\ref{eqn:L1L-1Hamiltonian}), $L_{1}+L_{-1}$ has also been investigated in a different context \cite{Rastelli:2001hh} in which the Hamiltonian was retained as $L_{0}+{\bar L}_{0}$. In this study, we  change the Hamiltonian itself to (\ref{eqn:L1L-1Hamiltonian}), and examine its consequences. In fact, this case turned out to correspond to the entanglement Hamiltonian as discussed in Ref. \cite{Wen:2016inm}, and further explored in Ref. \cite{Wen:2018vux,Wen:2018agb}, thus providing further motivation for the present research.

Let us elucidate the significance of the operator (\ref{eqn:L1L-1Hamiltonian}) in the context of  radial quantization \cite{Fubini:1972mf} and  dipolar quantization \cite{Ishibashi:2015jba,Ishibashi:2016bey}.
As is well-known, the $L_0, L_1$ and $L_{-1}$ operators constitute $sl(2,{\mathbb R})$ algebra. The combination of these operators,
\begin{equation}
x^{(0)}L_0+x^{(1)}L_1+x^{(-1)}L_{-1}, \label{eqn:linearcombi}
\end{equation}
can be mapped to 
\begin{equation}
x'^{(0)}L_0+x'^{(1)}L_1+x'^{(-1)}L_{-1},
\end{equation}
by the adjoint action of $sl(2,{\mathbb R})$; however, the following quadratic form of the coefficients, which is known as the quadratic Casimir element,
remains the same:
\begin{equation}
c^{(2)}\equiv \left(x^{(0)}\right)^2-4x^{(1)}x^{(-1)}= \left(x'^{(0)}\right)^2- 4x'^{(1)}x'^{(-1)} . \label{eqn:casimirdef}
\end{equation}

Using $c^{(2)}$,  the general linear combinations (\ref{eqn:linearcombi}) can be classified into three distinctive classes that are not accessible from each other by the $sl(2,{\mathbb R})$ action. Each class can be represented by a typical operator up to the overall rescaling: $L_{0}$ represents case $c^{(2)} >0 $ and $L_{0}-\frac12(L_{1}+L_{-1})$ represents  case $c^{(2)}=0$,  which correspond to  radial quantization and dipolar quantization, respectively. The final case, $c^{(2)}<0$, can be represented by $L_{1}+L_{-1}$, which signifies the importance of the operator in question.
Below, we investigate the $L_{1}+L_{-1}$ operator  by applying the formalism developed in \cite{Ishibashi:2015jba,Ishibashi:2016bey} and demonstrate that the three cases mentioned above, including $L_{1}+L_{-1}$, can be studied in a unified manner.

%
%
%
%

\section{Time-development vector field}

First, let us recapitulate the analysis in \cite{Ishibashi:2015jba,Ishibashi:2016bey}.
We introduce a set of differential operators, $\l_{\kappa}$, over the complex plane  with  label $\kappa$, in the following form:
\begin{equation}
\l_{\kappa}=-g(z)f_{\kappa}(z)\frac{\partial}{\partial z}, \label{eqn:Lkdef}
\end{equation}
where $g(z)$ and $f_{\kappa}(z)$ are  both holomorphic functions on the complex plane. Among $\l_{\kappa}$, we select the $\l_{0}$ operator and require
\begin{equation}
\l_{0} f_{\kappa} (z)= -\kappa f_{\kappa}(z),\label{eqn:fkdef}
\end{equation}
so that each $f_{\kappa}$ is an eigenfunction of $\l_{0}$ with the eigenvalue $-\kappa$. 
The solution to Eq. (\ref{eqn:fkdef}) is 
\begin{equation}
f_{\kappa}(z)=e^{\kappa \int dz/g(z)}, \label{eqn:fksol}
\end{equation}
which yields an simple expression for $\l_{0}$ as follows:
\begin{equation}
\l_{0}=-g(z)\frac{\partial}{\partial z}. \label{eqn:L0def}
\end{equation}
It can be seen that
\begin{equation}
f_{\kappa}(z)f_{\kappa'}(z)=e^{(\kappa+\kappa')\int^{z}\frac{dz}{g}}=f_{\kappa+\kappa'}(z), \label{eqn:ffprod}
\end{equation}
which enables the calculation of the commutation relations among $\l_{\kappa}$'s,
\begin{equation}
[\l_{\kappa}, \l_{\kappa'}]=(\kappa' -\kappa)g(z)f_{\kappa}(z)f_{\kappa'}(z)\frac{\partial}{\partial z} =
(\kappa -\kappa')
\l_{\kappa+\kappa'} \ . \label{eqn:Viranocent}
\end{equation}

It should be noted that 
the algebra generated by $\l_{\kappa}$ can be represented over the linear space spanned by $f_{\kappa}(z)$'s, since
\begin{equation}
\l_{\kappa}f_{\kappa'}(z)=f_{\kappa}(z)\l_{0}f_{\kappa'}(z)=-\kappa'f_{\kappa}(z)f_{\kappa'}(z)=-\kappa'f_{\kappa+\kappa'}(z).
\end{equation}
Because $\kappa$ is the index for the basis that spans the representation space,  
the representation theory may incur restrictions on $\kappa$. For example, if  $g(z)$ is set to be $z$,
\begin{equation}
f_{\kappa}=\exp(\kappa\int^{z}\frac{dz}{z})=z^{\kappa}, \label{eqn:fkord}
\end{equation}
up to possible $\kappa$ depending multipliers, which are omitted here.
It is natural for $f_{\kappa}=z^{\kappa}$ to be single-valued on the complex plane thus, each $\kappa$ should take an integer value
\footnote{ $\kappa$ can also  be a half-integer introducing a cut on the complex plane, which is useful for considering fermions on the world-sheet.}.
Another choice of $g(z)=z-\frac{z^{2}+1}{2}$ \cite{Ishibashi:2015jba,Ishibashi:2016bey}
leads to
\begin{equation}
f_{\kappa}=\exp(-\frac{2\kappa}{z-1}). \label{eqn:fkssd}
\end{equation}
In this case, $\kappa$ can  take any real number without inflicting multiple values on $f_{\kappa}(z)$.

The  procedure described above can be extended to the anti-holomorphic variable $\bar z$ by introducing,
\begin{equation}
{\bar \l}_{\kappa}\equiv -{\tilde g}({\bar z}){\tilde f}_{\kappa}(\bar z)\frac{\partial}{\partial {\bar z}}. \label{eqn:Lbkdf}
\end{equation}
The most natural choice for $\tilde g(\bar z)$ is the complex conjugate of $g(z)$ \footnote{It would be interesting to explore the ``heterotic'' case where $\tilde g({\bar z}) \neq g({\bar z})$.} , 
\begin{equation}
\tilde g({\bar z}) \equiv \overline{g(z)}= g({\bar z}). 
\end{equation}
It then follows that 
\begin{equation}
\tilde f_{\kappa}({\bar z})=f_{\kappa}({\bar z}).
\end{equation}
$\bar\l_{\kappa}$ satisfies the following commutation relation:
\begin{equation}
[\bar \l_{\kappa}, \bar \l_{\kappa'}]=
(\kappa -\kappa')
\bar \l_{\kappa+\kappa'} ,  \label{eqn:Virabarnocent}
\end{equation}
which is isomorphic to Eq. (\ref{eqn:Viranocent}).

We can define the time-translation operator, using Eq. (\ref{eqn:L0def}) and
\begin{equation}
{\bar \l}_{0}=-g({\bar z})\frac{\partial}{\partial {\bar z}}, \label{eqn:L0bdf}
\end{equation}
through the following expression:
\begin{equation}
-\frac{\partial}{\partial t} \equiv \l_{0} +{\bar  \l}_{0},
\end{equation}
which  yields the time coordinate, $t$. The space coordinate, $s$, can  be obtained by the following orthogonal operator:
\begin{equation}
-\frac{\partial}{\partial s} \equiv i \left( \l_{0} -{\bar  \l}_{0}\right).
\end{equation}
The above two equations can be summarized in the following explicit matrix form:
\begin{equation}
\left(\begin{array}{c}\frac{\partial}{\partial t}  \\[5pt] \frac{\partial}{\partial s}\end{array}\right) =\left(\begin{array}{cc}g(z) & g({\bar z}) \\[5pt]ig(z) & -ig({\bar z})\end{array}\right)\left(\begin{array}{c}\frac{\partial}{\partial z} \\[5pt] \frac{\partial}{\partial {\bar z}}\end{array}\right). \label{eqn:jacobitszzbar}
\end{equation}
Then, the relation between the coordinates $t, s$ and $z, \bar z$ can be expressed succinctly \cite{Ishibashi:2015jba,Ishibashi:2016bey} as
follows:
\begin{equation}
t+is=\int^z \frac{dz}{g(z)}. \label{eqn:intzgz}
\end{equation}

We 
now investigate the effect of selecting Eq. (\ref{eqn:L1L-1Hamiltonian}) as the time-development operator of the system using the aforementioned formalism. The choice of $g(z)$ that corresponds to Eq. (\ref{eqn:L1L-1Hamiltonian}) should be 
\begin{equation}
g(z)=z^{2}+1. \label{eqn:gzz2p1}
\end{equation}
The time translation on $z$ is then expressed as
\begin{equation}
-\frac{\partial}{\partial t}=-z^{2}\frac{\partial}{\partial { z}} -\frac{\partial}{\partial { z}} -{\bar z}^{2}\frac{\partial}{\partial {\bar z}}-\frac{\partial}{\partial {\bar z}}. \label{eqn:l1l-1vec}
\end{equation}
The relation between the above vector field (\ref{eqn:l1l-1vec}) (or the choice of $g(z)$) and the operator (\ref{eqn:L1L-1Hamiltonian}) can be easily deduced 
by noting the expression of the Virasoro generators as follows:
\begin{equation}
L_{n}=\frac{1}{2\pi i} \oint dz z^{n+1} T(z).
\end{equation}

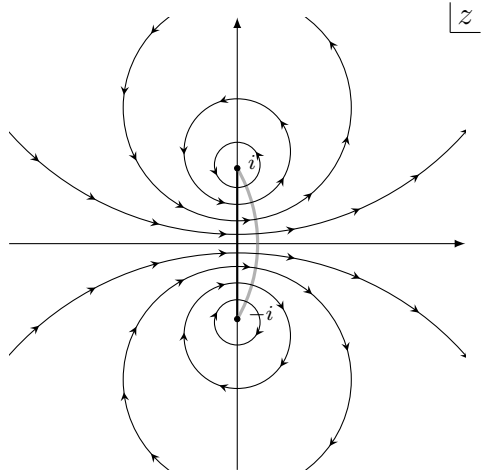
\begin{figure}[tbh]
\centering
\begin{tikzpicture}[>=latex,decoration={
markings,mark=between positions 0.1 and 0.97 step 0.9cm with {\arrow{stealth}}}]
\node at (3,3) {$z$};
\draw (3.2,2.8) -- (2.8,2.8);
\draw (2.8,2.8) -- (2.8,3.2);
  \draw[->] (-3,0) -- (3,0);
  \draw[->] (0,-3) -- (0,3);
  \draw[thick] (0,-1) -- (0,1);
\draw[very thick,gray, opacity=0.6] (0,-1) arc (-30:30:2cm);
  \filldraw (0,1) circle (1pt) node[anchor=west] {$^{i} $};
  \filldraw (0,-1) circle (1pt)node[anchor=west] {$^{-i}$};

 \clip (-3,-3) rectangle (3,3);
\foreach \R in {0.3,0.7,1.5,4} 
{
  \draw [postaction={decorate}](0, {sqrt(\R*\R+1)}) circle (\R cm);
 \draw [postaction={decorate}](\R,-{sqrt(\R*\R+1)}) arc (360:0:\R cm);
 }
  \end{tikzpicture}
\caption{Flow of time $t$ generated by $\left(z^{2}+1\right)\frac{\partial}{\partial { z}} +\left({\bar z}^{2}+1\right)\frac{\partial}{\partial {\bar z}}$. $t=-\pi/2 $ (or $\pi/2$) on the solid line between $i$ and $-i$, and $t=0$ on  the remainder of the imaginary axis. The value of $t$ is unavoidably periodic. 
A line with constant $t$ is shown in gray.
} \label{fig:timeflow}
\end{figure}

%
%
The selection (\ref{eqn:gzz2p1}) yields \cite{Ishibashi:2016bey}
\begin{equation}
f_{\kappa}=\exp(\kappa\int^{z}\frac{dz}{z^{2}+1})e^{\kappa \cdot const.}=\left(\frac{i+z}{i-z}\right)^{-\frac{i}{2}\kappa}=e^{\kappa \arctan z}
,
\label{eqn:fk1-1}
\end{equation}
and, finally,
\begin{equation}
\l_{\kappa}=-(z^{2}+1)e^{\kappa \arctan z}\frac{\partial}{\partial z}.
\end{equation}
Possible constant multiplications are omitted for the sake of brevity.

A notable idiosyncrasy of the solution (\ref{eqn:fk1-1}) is the appearance of the multivalued function $
\kappa\arctan z$, which may yield  ambiguous multiplicative factors 
\begin{equation}
e^{\pi n \kappa}  \ \ \ \ \ \  n \in \mathbb{Z},
\end{equation}
without a proper specification of the principal value.
The reason for these multiple values is clear from Fig. \ref{fig:timeflow}. With the proper selection of the principal value for $\arctan$, $t=-\frac{\pi}{2}$ at the thick line between $z=i$ and $z=-i$ as illustrated in Fig. \ref{fig:timeflow}. Along the time flow, however, $t$ becomes $\frac{\pi}{2}$ after encircling either $z=i$ or $z=-i$. As $t$ develops further, it returns to the same point on the $z$ coordinate with different values for $t$.
To be more precise, the time and space coordinates can be given by Eq. (\ref{eqn:intzgz}) as follows:
\begin{equation}
t+is=\int^{z}\frac{dz}{z^{2}+1}=\frac{i}{2}\ln\left(\frac{z+i}{z-i}\right). \label{eqn:tsz21}
\end{equation}
If we convert the argument of the logarithm in Eq. (\ref{eqn:tsz21})
to polar coordinates 
\begin{equation}
\frac{z+i}{z-i}\equiv Re^{i\theta},
\end{equation}
it is then simple to discern the time and space coordinates in terms of $R$ and $\theta$ as follows:
\begin{equation}
\frac{i}{2}\ln\left(\frac{z+i}{z-i}\right)=-\frac12\theta+\frac{i}{2}\ln R=t+is.
\end{equation}

%
%
%
%
%
%
%
%
%

\section{Conserved charges and the Virasoro algebra}
The analysis above produces a set of (conformal) Killing vectors,
\begin{equation}
g(z)f_{\kappa}(z) ,
\end{equation}
in the complex notation.
We can now define the conserved charges by integrating the Noether current, which is the product of the energy momentum tensor and the Killing vector:
\begin{equation}
\L_{\kappa} \equiv \frac{1}{2\pi i}\int_{\cal C} dz g(z)f_{\kappa}(z) T(z). \label{eqn:defLkappa}
\end{equation}
Here, $\cal C$ denotes a contour on  which $t$ is constant and $s$ takes all  possible values. An example of such a contour is depicted as a gray line in Fig. \ref{fig:timeflow}.
The definition of the anti-holomorphic charge, $\bar\L_{\kappa}$, should be trivial. It should be noted that $\L_{0}$ and $\bar\L_{0}$ are significant charges among others because
\begin{equation}
\L_{0}=L_{1}+L_{-1} , \ \ \bar \L_{0}=\bar L_{1}+\bar L_{-1} ,
\end{equation}
for the selection of $g(z)$, as in Eq. (\ref{eqn:gzz2p1}).

The operator product expansion of the energy momentum tensor is governed by  conformal symmetry on the $z$-plane and takes the following form:
\begin{equation}
 T(z)T(z') \sim \frac{\cc/2}{(z-z')^{4}}+\frac{2T(z')}{(z-z')^{2}} + \frac{\partial_{z'}T(z')}{z-z'} + \cdots ,
\end{equation}
where $\cc$ is the central charge of CFT in question. Then, the commutation relations among the conserved charges $\L_{\kappa}$'s lead to the following integration:
\footnote{Strictly speaking, the integral near the singularity should be carefully examined as has been done in \cite{Kishimoto:2018ekq}, especially in the appendix B.}
\begin{eqnarray}
[\L_{\kappa} , \L_{\kappa'} ]&=&\frac{1}{2\pi i}   \int_{\cal C}  dz' g(z')f_{\kappa'}(z')\int_{\Czzdash} 
\frac{dz}{2\pi i}
g(z)f_{\kappa}(z) \mathbf{T}\left(T(z) T(z')  \right) \nonumber \\
&=& \frac{1}{2\pi i}   \int_{\cal C} dz' g(z')f_{\kappa'}(z')\int_{\Czzdash} 
\frac{dz}{2\pi i}
g(z )f_{\kappa}(z )  \label{eqn:LLcom}\\
&& \times \left(  \frac{\cc/2}{(z -z')^{4}}+\frac{2T(z')}{(z-z')^{2}} + \frac{\partial_{z'}T(z')}{z-z'} \right).\nonumber 
\end{eqnarray}
Performing the contour integral around $z'$  in Eq. (\ref{eqn:LLcom}) yields
\begin{eqnarray}
[\L_{\kappa} , \L_{\kappa'} ]&=& \frac{\cc}{12}    \int_{\cal C}  \frac{dz' }{2\pi i}
\left\{ g \frac{\partial^{3}g}{\partial z'^{3}}+
\kappa\left( 2\frac{\partial^{2}g}{\partial z'^{2}}-\frac{1}{g}\left(\frac{\partial g}{\partial z'}\right)^{2}\right)+\frac{\kappa^{3}}{g}\right\}f_{\kappa+\kappa'}(z') \nonumber \\
&&+ (\kappa-\kappa')    \int_{\cal C}  \frac{dz' }{2\pi i} g(z')f_{\kappa+\kappa'}(z')T(z'), \label{eqn:elkelkd}
\end{eqnarray}
where the last term of the righthand side is nothing but
\begin{equation}
(\kappa-\kappa')  \L_{\kappa+\kappa'}. \nonumber
\end{equation}
If we denote the integral part of the central extension in the first term of the righthand side of Eq. (\ref{eqn:elkelkd}) as
\begin{equation}
\mbox{CI}[\kappa|\kappa'] \equiv  \int_{\cal C}  \frac{dz }{2\pi i}
\left\{ g \frac{\partial^{3}g}{\partial z^{3}}+
\kappa\left( 2\frac{\partial^{2}g}{\partial z^{2}}-\frac{1}{g}\left(\frac{\partial g}{\partial z}\right)^{2}\right)+\frac{\kappa^{3}}{g}\right\}f_{\kappa+\kappa'}(z),
\label{eqn:CIdef}
\end{equation}
the commutation relations read
\begin{equation}
[\L_{\kappa} , \L_{\kappa'} ]=(\kappa-\kappa')  \L_{\kappa+\kappa'}+\frac{\cc}{12}\mbox{CI}[\kappa|\kappa'] .    \label{VirCI}
\end{equation}
For the charges, $\L_{\kappa}$, to satisfy the Virasoro algebra in Eq. (\ref{VirCI}), CI$[\kappa|\kappa']$ must vanish unless $\kappa+\kappa'=0$. Otherwise, a certain part of the Jacobi identity is breached (see, for example,\cite{Blumenhagen:2009zz}). In the following, we evaluate CI$[\kappa|\kappa']$ explicitly to verify whether CI$[\kappa|\kappa']$ can be zero for $\kappa+\kappa'\neq0$.

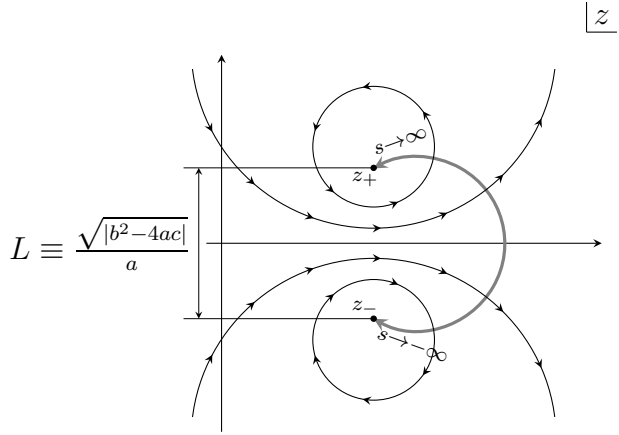
\begin{figure}[tbh]
\centering
\begin{tikzpicture}[>=stealth,decoration={
markings,mark=between positions 0.1 and 0.97 step 0.9cm with {\arrow{stealth}}}]
\node at (3,3) {$z$};
\draw (3.2,2.8) -- (2.8,2.8);
\draw (2.8,2.8) -- (2.8,3.2);
  \draw[->] (-2.2,0) -- (3,0);
  \draw[->] (-2,-2.5) -- (-2,2.5);
  \draw[very thin] (0,1) -- (-2.5,1);
    \draw[very thin] (0,-1) -- (-2.5,-1);
    \draw[very thin,<->] (-2.3,1)--(-2.3,-1);
    \node at (-3.6,0) {$L\equiv{\frac{\sqrt{|b^{2}-4ac|}}{a}}$};
\draw[gray, very thick,<->] (0,-1) arc (-120:120:1.15cm);
\node[anchor=west,rotate=20] at (-0.1,1.1) {$^{s\rightarrow \infty}$};
\node[anchor=west,rotate=-20] at (-0.1,-1.2) {$^{s\rightarrow -\infty}$};
  \filldraw (0,1) circle (1pt);
  \node at (-0.1,0.8) {$^{z_{+}} $};
  \filldraw (0,-1) circle (1pt);
  \node at (-0.1,-0.9) {$^{z_{-}}$};

 \clip (-3,-2.3) rectangle (3,2.3);
\foreach \R in {0.8,2.4} 
{
  \draw [postaction={decorate}](0, {sqrt(\R*\R+1)}) circle (\R cm);
 \draw [postaction={decorate}](\R,-{sqrt(\R*\R+1)}) arc (360:0:\R cm);
 }
  \end{tikzpicture}
\caption{Mapping of time $t$ and space $s$ onto $z$, generated by $g(z)=az^{2}+bz+c\equiv a(z-z_{+})(z-z_{-})$. Solid lines with arrows represent the flow of $t$. 
The gray line connecting $z_{\pm}$ is a possible contour, $\cal C$, where $t$ is constant and $s$ changes; in particular, $s\rightarrow \pm \infty$ near $z_{\pm}$, respectively.} \label{fig:generalgflow}
\end{figure}

Evaluating the value of CI$[\kappa|\kappa']$ involves function $g$. Function $g$ is explicitly given by Eq. (\ref{eqn:gzz2p1}) for the case at hand; however, it is useful to consider a more general case,
\begin{equation}
g(z)=az^{2}+bz+c, \label{eqn:geng}
\end{equation}
where we assume
\begin{equation}
b^{2}-4ac <0 \ , 
\end{equation}
to keep the quadratic Casimir element $c^{(2)}$ negative (see Eq. (\ref{eqn:casimirdef})). We also limit  $a$ to be positive for the sake of  notational simplicity. 
The flow generated by $g(z)$ in Eq. (\ref{eqn:geng}) is illustrated in Fig. \ref{fig:generalgflow}.

The terms inside the braces in the definition of CI (\ref{eqn:CIdef}) can be easily demonstrated to amount to
\begin{equation}
g \frac{\partial^{3}g}{\partial z^{3}}+
\kappa\left( 2\frac{\partial^{2}g}{\partial z^{2}}-\frac{1}{g}\left(\frac{\partial g}{\partial z}\right)^{2}\right)+\frac{\kappa^{3}}{g}
=
\frac{1}{g(z)} 
\left(-(b^{2}-4ac)\kappa+\kappa^{3}\right),
\end{equation}
thus yielding
\begin{equation}
\mbox{CI}[\kappa|\kappa']= \left(-(b^{2}-4ac)\kappa+\kappa^{3}\right) \int_{\cal C}  \frac{dz }{2\pi i} \frac{f_{\kappa+\kappa'}(z)}{g(z)}. \label{eqn:CIcalc}
\end{equation}
Noting that differentiating Eq. (\ref{eqn:fksol}) yields
\begin{equation}
df_{\kappa}=\kappa \frac{dz}{g(z)} f_{\kappa},
\end{equation}
the integral in Eq. (\ref{eqn:CIcalc}) can be further simplified as
\begin{equation}
\int_{\cal C}  \frac{dz }{2\pi i} \frac{f_{\kappa+\kappa'}(z)}{g(z)}=\int_{\cal C}  \frac{df_{\kappa+\kappa'} }{2\pi i (\kappa+\kappa')} =\left. \frac{f_{\kappa+\kappa'} }{2\pi i (\kappa+\kappa')} \right|_{\partial{\cal C}} \ ,
\end{equation}
for  $\kappa+\kappa' \neq 0$.
If we denote two roots of $g(z)=0$ as $z_{\pm}$, whose imaginary parts are positive and negative, respectively, they constitute the boundary of $\cal C$. Thus,
\begin{equation}
\left. \frac{f_{\kappa+\kappa'} }{2\pi i (\kappa+\kappa')} \right|_{\partial{\cal C}}=\frac{1}{2\pi i (\kappa+\kappa')} \left( f_{\kappa+\kappa'}(z_{+})-f_{\kappa+\kappa'}(z_{-})\right) .
\label{eqn:fzp-fzm}\end{equation}

It should be noted that expression (\ref{eqn:fk1-1}) 
is generalized as follows:
\begin{equation}
f_{\kappa}(z)=\exp\left(\kappa\int^{z}\frac{dz}{a(z-z_{+})(z-z_{-})}\right)=\exp\left(
\frac{\kappa}{a(z_{+}-z_{-})}\ln\left(\frac{z-z_{+}}{z-z_{-}}\right)\right) ,   \label{eqn:fkappaexp}
\end{equation}
where possible multiplicative constants are neglected.
$f_{\kappa+\kappa'}(z_{+})$ and $f_{\kappa+\kappa'}(z_{-})$  are apparently divergent; as a result, the evaluation of Eq. (\ref{eqn:fzp-fzm}) is non-trivial. 
Therefore, we introduce the cut-off, $\varepsilon$, near the fixed points $z=z_{\pm}$, bearing in mind the application to entanglement entropy (Fig. \ref{fig:cutoff}). 
\begin{figure}[tbh]
\centering
\begin{tikzpicture}[>=stealth,decoration={
markings,mark=between positions 0.1 and 0.97 step 0.9cm with {\arrow{stealth}}}]
\filldraw[white] (0,0) circle (12pt);
\filldraw[gray, opacity=0.3] (0,0) circle (11.5pt);
\draw [dashed, thick] (0,0) circle (12pt);
\draw[thin, <->] (0,0) -- (0.3,0.3) node[anchor=north]{$^{\varepsilon}$};
 \filldraw (0,0) circle (1pt);
 \node at (-0.12,-0.12) {$z_{+}$};
 \clip (-2,-2) rectangle (2,2.3);
\foreach \R in {0.65,1.05,1.45, 1.85} 
{
  \draw [postaction={decorate}](0, 0) circle (\R cm);
 }
 \draw[gray, very thick,<->] (6.3,2) arc (60:149.5:4.5cm);
 \node[anchor=west,rotate=33] at (1,1.6) {$^{s\rightarrow \infty}$};
  \end{tikzpicture}
\caption{Cut-off region near $z_{+}$ (colored in gray). The radius of the cut-off is $\varepsilon$ in the $z$ plane. 
} \label{fig:cutoff}
\end{figure}
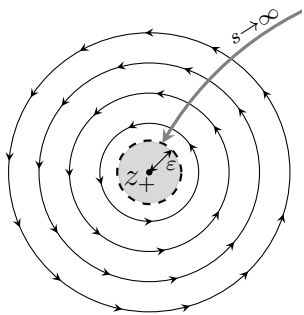

\begin{figure}[tbh]
\centering
\begin{tikzpicture}[>=latex,decoration={
markings,mark=between positions 0.1 and 0.99 step 0.5cm with {\arrow{stealth}}}]
\draw[dashed] (-3,-3) --(-3,-2.5);
\draw (-3, -2.5) -- (-3,2.2);
\draw[dashed] (-3,2.2) --(-3,3.2);
\draw[->] (-3, 3.2) -- (-3,3.5) node[anchor=west] {$s$};
\draw[->] (-2.5, -3.6) -- (2.5, -3.6) node[anchor=south] {$t$};
\draw[very thick] (2, -2) -- (2, 2);
\draw (-3.05,2) -- (-2.95,2) node [anchor=east]{$\frac{1}{aL}\ln(\frac{L}{\varepsilon})$};
\draw[very thick] (-2, -2) -- (-2, 2);
\draw (-3.05,-2) -- (-2.95,-2) node [anchor=east]{-$\frac{1}{aL}\ln(\frac{L}{\varepsilon})$};
\draw[thick,dashed] (-2,2) --(2, 2);
\draw[thick,dashed] (-2,-2) --(2, -2);
\draw(-2,-3.65) -- (-2, -3.55) node [anchor=south]{$-\frac{\pi}{aL}$};
\draw(2,-3.65) -- (2, -3.55) node [anchor=south]{$\frac{\pi}{aL}$};
\draw[gray, opacity=0.3,very thick, ->] (0,2.9) -- (0,3.4);
\node [gray, opacity=0.7] (A) at (0,3.5) {$^{z=z_{+}}$};
\draw[gray, opacity=0.3,very thick, ->] (0,-2.9) -- (0,-3.4);
\node [gray, opacity=0.7] (A) at (0,-3.5) {$^{z=z_{-}}$};
\filldraw[gray, opacity=0.3] (-2,2) -- (-2,2.5) -- (2,2.5) -- (2,2);
\filldraw[gray, opacity=0.3]  (-2,-2) -- (-2,-2.5) -- (2,-2.5) -- (2,-2);

\foreach \A in {0.05,0.2}
{
\filldraw[gray, opacity=0.3] (-2,2.5+\A) -- (-2,2.6+\A) -- (2,2.6+\A) -- (2,2.5+\A);
\filldraw[gray, opacity=0.3]  (-2,-2.5-\A) -- (-2,-2.6-\A) -- (2,-2.6-\A) -- (2,-2.5-\A);
}

\foreach \S in {1.5,1,0.5, 0,-0.5,-1,-1.5}
{\draw [postaction={decorate}](-2,\S) -- (2, \S);
}
%
  \end{tikzpicture}
\caption{Time translation in $s-t$ coordinates. The cut-off is represented by the dashed line adjacent to the gray-colored area, where the fixed points, $z_{+}$ or $z_{-}$ in the $z$-plane, are located infinitely far away. The circles with radius $\varepsilon$ in the $z$-plane, and the cut-off in the $s$ coordinate  is located at $s=\pm \frac{1}{aL} \ln(\frac{L}{\varepsilon})$. 
} \label{fig:tsgraph}
\end{figure}
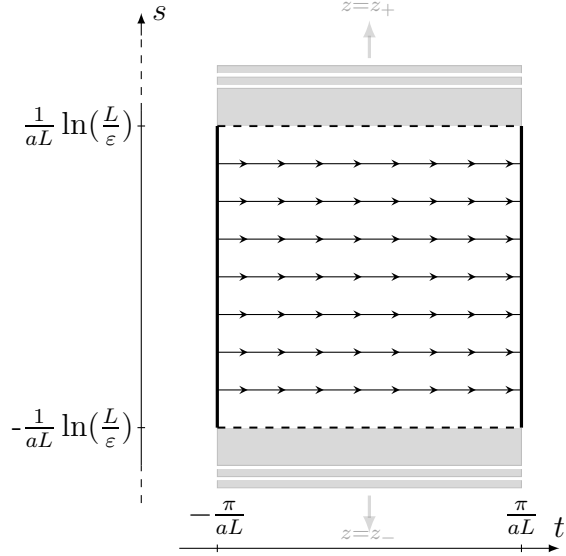

The structure near $z_{\pm}$ in terms of $t$ and $s$ can be determined from the following generalization of Eq. (\ref{eqn:tsz21}):
\begin{equation}
t+is=\int^{z}\frac{dz}{az^{2}+bz+c}=\frac{1}{a(z_{+}-z_{-})}\ln\left(\frac{z-z_{+}}{z-z_{-}}\right).
\end{equation}
Introducing $L$ as the length of the separation between $z_{+}$ and $z_{-}$ in the imaginary direction,
\begin{equation}
z_{+}-z_{-}=i\frac{\sqrt{|b^{2}-4ac|}}{a}\equiv iL,
\end{equation}
and setting $z$ as the cut-off boundary 
$z= z_{\pm}+\varepsilon e^{i\theta_\varepsilon}$ in the above equation, we obtain

%
%
\begin{equation}
t+is \sim \frac{\pm \theta_\varepsilon-\frac{\pi}{2}}{aL}\pm i\frac{1}{aL} \cdot \ln \left(\frac{L}{\varepsilon}\right)  \quad \mbox{at}  \, z\sim z_{\pm} \quad \mbox{and} \ \ \varepsilon \sim 0.
\end{equation}
Thus, in terms of $t$ and $s$, the cut-off boundaries are located at
\begin{equation}
z=z_{\pm}+e^{\mp aLs}e^{\pm i(aLt+\pi/2)} ,
\end{equation}
respectively; Fig. \ref{fig:tsgraph} provides a depiction.
Equation (\ref{eqn:fzp-fzm}) can then be evaluated using the cut-off, and the expression of CI$[\kappa|\kappa']$ for $\kappa+\kappa'\neq0$ can be obtained as follows:
\begin{eqnarray}
\mbox{CI}[\kappa|\kappa']_{\kappa+\kappa'\neq0}&= &
\left(-(b^{2}-4ac)\kappa+\kappa^{3}\right)\frac{e^{
%
\left(\kappa+\kappa'\right)t
}}{2\pi i (\kappa+\kappa')} \nonumber \\&&\times 
\left[ e^{i \frac{\kappa+\kappa'}{aL}\ln\left(\frac{L}{\varepsilon}\right)}- e^{-i \frac{\kappa+\kappa'}{aL}\ln\left(\frac{L}{\varepsilon}\right)} \right]. 
\label{eqn:CIkknonzero}
\end{eqnarray}
Finally, we uncover that  there should be an integer $n$ so that
\begin{equation}
2i\frac{\kappa+\kappa'}{aL}\ln\left(\frac{L}{\varepsilon}\right)=2\pi i n, \label{eqn:ccc2v}
\end{equation}
for Eq. (\ref{eqn:CIkknonzero}) to vanish.

We could require that $\kappa$, which was originally introduced as the label for the (differential) operator, takes the following values:
\begin{equation}
\kappa=\frac{\pi aL}{\ln\left({L}/{\varepsilon}\right)}n, \ \ \ n\in\mathbb Z \ \mbox{or}  \ \mathbb Z +\frac12   \    . \label{eqn:kappaLn}
\end{equation}
However, we rather homogeneously rescale the parameters $a,b,c$ in $g(z)$, which governs the time development,
\begin{equation}
a \rightarrow \xi a , \ b \rightarrow \xi b , \ c \rightarrow \xi c , \ {i.e.} \ g(z) \rightarrow \xi g(z),  \label{eqn:rescale}
\end{equation}
and demand
\begin{equation}
\frac{\pi aL}{\ln\left({L}/{\varepsilon}\right)}=1   \label{eqn:lambdacond}
\end{equation}
so that $\kappa$ can be either an integer or half-integer.
Noting that $L$ is invariant under the rescaling (\ref{eqn:rescale}), it is evident that selecting $\xi$ as 
\begin{equation}
\xi=\frac{\ln \left(\frac{L}{\varepsilon}\right)}{a\pi L} \label{eqn:xidef}
\end{equation}
satisfies Eq. (\ref{eqn:lambdacond}).

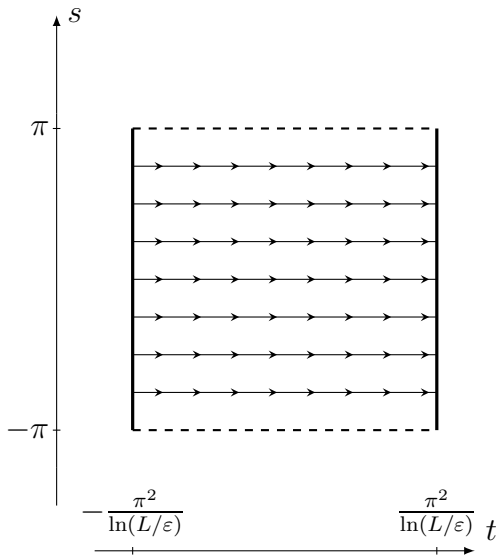
\begin{figure}[tbh]
\centering
\begin{tikzpicture}[>=latex,decoration={
markings,mark=between positions 0.1 and 0.99 step 0.5cm with {\arrow{stealth}}}]
\draw (-3,-3) --(-3,-2.5);
\draw (-3, -2.5) -- (-3,2.2);
\draw(-3,2.2) --(-3,3.2);
\draw[->] (-3, 3.2) -- (-3,3.5) node[anchor=west] {$s$};
\draw[->] (-2.5, -3.6) -- (2.5, -3.6) node[anchor=south west] {$t$};
\draw[very thick] (2, -2) -- (2, 2);
\draw (-3.05,2) -- (-2.95,2) node [anchor=east]{$\pi$};
\draw[very thick] (-2, -2) -- (-2, 2);
\draw (-3.05,-2) -- (-2.95,-2) node [anchor=east]{$-\pi$};
\draw[thick,dashed] (-2,2) --(2, 2);
\draw[thick,dashed] (-2,-2) --(2, -2);
\draw(-2,-3.65) -- (-2, -3.55) node [anchor=south]{$-\frac{\pi^{2}}{\ln\left({L}/{\varepsilon}\right)}$};
\draw(2,-3.65) -- (2, -3.55) node [anchor=south]{$\frac{\pi^{2}}{\ln\left({L}/{\varepsilon}\right)}$};
%

\foreach \S in {1.5,1,0.5, 0,-0.5,-1,-1.5}
{\draw [postaction={decorate}](-2,\S) -- (2, \S);
}
%
  \end{tikzpicture}
\caption{Time translation in terms of  rescaled $g(z)$. By  rescaling, the functional values coincide at the upper and  lower dashed lines; the dashed lines can thus be identified. The thick lines on the right and  left is actually the same line in the $z$-plane; thus, it is virtually a torus geometry.  %
} \label{fig:tsrescaled}
\end{figure}

The rescaling (\ref{eqn:rescale}) also affects the range of $t$ and $s$ as
\begin{equation}
-\pi<s<\pi , \ \  \, -\frac{\pi^{2}}{\ln\left(\frac{L}{\varepsilon}\right)} <t <\frac{\pi^{2}}{\ln\left(\frac{L}{\varepsilon}\right)} \ . \label{eqn:stperiod}
\end{equation}
Thus, there is now a torus with the moduli parameter,
\begin{equation}
\tau=i\frac{\pi}{\ln\left({L}/{\varepsilon}\right)}, \label{eqn:modulidef}
\end{equation}
on which the path integral should be performed, as depicted in Fig. \ref{fig:tsrescaled}.

The introduction of  cut-off $\varepsilon$ and the corresponding rescaling of $g(z)$ eliminate the undesirable contribution to $\mbox{CI}[\kappa|\kappa']$ for $\kappa\neq-\kappa'$. However, the case for $\kappa=-\kappa'$ yields 
\begin{equation}
\mbox{CI}[\kappa|-\kappa]=\left(\kappa^{3}-(b^{2}-4ac)\kappa\right) \int_{\cal C_{\varepsilon}}  \frac{dz }{2\pi i} \frac{1}{g(z)}, 
\end{equation}
by invoking Eqs. (\ref{eqn:fksol}) and (\ref{eqn:CIcalc}). Here we denote the contour with the cut-off as $C_{\varepsilon}$.
The above expression can be further evaluated as
\begin{equation}
\mbox{CI}[\kappa|-\kappa]=\left(\kappa^{3}-(b^{2}-4ac)\kappa\right)\frac{\ln\frac{L}{\varepsilon}+\frac{i\pi}{2}}{a\pi L}. 
\end{equation}
Employing the divergent rescaling (\ref{eqn:rescale}) and neglecting the finite $i\pi/2$ term,
\begin{equation}
\mbox{CI}[\kappa|-\kappa]=\left(\kappa^{3}-\xi^{2}(b^{2}-4ac)\kappa\right). \label{eqn:cikmk}
\end{equation}
Thus, we arrive at the following Virasoro algebra:
\begin{equation}
[\L_{\kappa} , \L_{\kappa'} ]=(\kappa-\kappa')  \L_{\kappa+\kappa'}+\frac{\cc}{12}\left(\kappa^{3}-\xi^{2}(b^{2}-4ac)\kappa\right)\delta_{\kappa,-\kappa'} ,    \label{Virxi}
\end{equation}
where $\kappa$ is either an integer or half-integer. In addition, $a, b, $ and $c$ are the original values before rescaling, as introduced in Eq. (\ref{eqn:geng}). However, $\L_{\kappa}$ is defined by Eq. (\ref{eqn:defLkappa}) with the rescaled $\xi g(z)$ and $f_{\kappa}(z)$ which are also defined by the rescaled $\xi g(z)$ in Eq. (\ref{eqn:fkdef}). 

Thus, we have established the Virasoro algebra with the exception of the divergent term in the central charge. However, this central extension term that is proportional to $\kappa\delta_{\kappa, -\kappa'}$ can be absorbed into the constant shift in $\L_{0}$ as follows:
\begin{equation}
\L_{0} \rightarrow \L_{0} -\frac{\cc}{24}\xi^{2}(b^{2}-4ac). \label{eqn:L0shift}
\end{equation}
A similar procedure is performed when one considers the CFT on a cylinder by  mapping from a complex plane to a cylinder. In this case, however, the shift is $\frac{\cc}{24}$. The difference is that the entire complex plane or the Riemann sphere is mapped into a cylinder, whereas in our case, the degrees of freedom in the two disks with radius $\varepsilon$ are discarded. We simply take  advantage of the coincidence of the values of $f_{\kappa}$ and interpret the space as a torus. 
However,  once the shift  in Eq. (\ref{eqn:L0shift}) is performed, the resulting set of Virasoro generators $\L_{\kappa}$ yields the same central extension term as the one on the torus. 

The procedure performed up to now is recapitulated below. We began with the time foliation of the Riemann sphere governed by $g(z)=az^{2}+bz+c$, where parameters $a,b$ and $c$ satisfy $b^{2}-4ac \leq 0$. The selection of the parameters is exemplified by the case $a=c=1$ and $b=0$, which corresponds to the $L_{1}+L_{-1}$ operator. The flow of time exhibits two fixed points separated by $L$ as depicted in Fig. \ref{fig:timeflow2}. Along this time development, we can define a set of conserved charges as Eq. (\ref{eqn:defLkappa}). Requiring these conserved charges to form a Virasoro algebra, we are led to introduce the cut-off around the two fixed points (Fig. \ref{fig:cutoff}) and the periodic (or anti-periodic) boundary condition by means of rescaling (\ref{eqn:rescale}).
Now, we have  the CFT on the torus with central charge $c$ and modular parameter $\tau $, as defined in Eq. (\ref{eqn:modulidef}).
On the torus, the time development is invoked by $\L_{0}+\bar\L_{0}$, which is shifted from the original definition (\ref{eqn:defLkappa}) by Eq. (\ref{eqn:L0shift}). The  procedure  described above is connected with entanglement entropy as explained in the next section\footnote{For the review and the background of the topic, we refer \cite{1609.00026} and references therein. }.


\section{Entanglement Entropy
}

Consider a function  defined on the line between the two fixed points, which we refer to as $L$ (Fig. \ref{fig:timeflow2}):
\begin{equation}
\phi^L_{i}(x),
\end{equation}
where subscript $i$ denotes the index for the basis that spans  the Hilbert space of such functions. Thus,
$\phi^L_{i}(x)$  can also be  considered a state vector of the Hilbert space that corresponds to space $L$,
\begin{equation}
|\phi^L_{i}\rangle  \leftrightarrow \phi^L_{i}(x).
\end{equation}
The time-dependent state is constructed by applying the time development as
\begin{equation}
|\phi_{i}^{L'}(t)\rangle \equiv e^{-tH}|\phi_{i}^L\rangle,
\end{equation}
where $H$ is the generator of the time development, namely the Hamiltonian, and $L'$ is the region developed from the original segment $L$ during time period $t$.

\begin{figure}[tbh]
\centering
\begin{tikzpicture}[>=latex,decoration={
markings,mark=between positions 0.1 and 0.97 step 0.9cm with {\arrow{stealth}}}]
\node at (3,3) {$z$};
\draw (3.2,2.8) -- (2.8,2.8);
\draw (2.8,2.8) -- (2.8,3.2);
\draw[very thick] (0,1) -- (0,-1);
\draw[thick, dashed](0,1)--(0,3.1);
\draw[thick, dashed](0,-1)--(0,-3.1);
  \filldraw[gray] (0,1) circle (2pt) ;
  \filldraw [gray] (0,-1) circle (2pt);
  \draw[very thin] (0,1) -- (-1,1);
    \draw[very thin] (0,-1) -- (-1,-1);
    \draw[very thin,<->] (-0.6,1)--(-0.6,-1);
    \node at (-0.8,0) {$L$};
        \node at (0.16,2.5) {$L^c$};
            \node at (0.16,-2.5) {$L^c$};
 \clip (-3,-3) rectangle (3,3);
\foreach \R in {0.3,0.7,1.5,4} 
{
  \draw [postaction={decorate}](0, {sqrt(\R*\R+1)}) circle (\R cm);
 \draw [postaction={decorate}](\R,-{sqrt(\R*\R+1)}) arc (360:0:\R cm);
 }
  \end{tikzpicture}
\caption{The flow of time $t$ can be considered to begin at a section of space with length $L$ (solid line). We refer to this section of space as $L$ without fear of confusion. The rest of the space, including  infinity, is depicted as a dashed line and denoted  $L^c$. After sweeping the entire complex plane, including  infinity, with the exception of  the cut-off region (gray dots), the flow returns to the original line.
} \label{fig:timeflow2}
\end{figure}
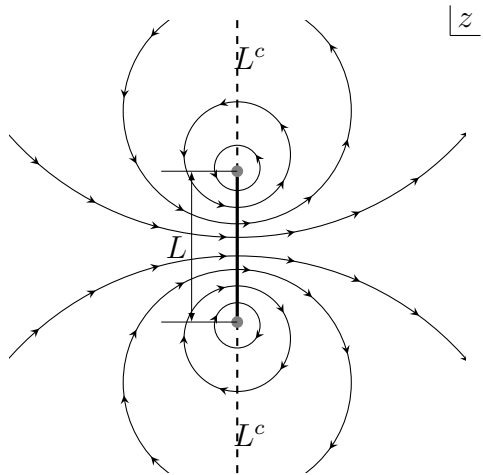

\begin{figure}[htbp]
\begin{center}
\begin{tikzpicture}[scale=0.5, every node/.style={scale=0.65},rotate=270,decoration={
markings,
mark=between positions 0.1 and 0.9 step 8mm with {\arrowreversed{stealth}}} ]
\def\R{3} 
   \draw (0,0) circle  (\R cm and \R cm);
 \shade[ball color=blue!5!white,opacity=0.4] (-\R,0)   (0,0) circle  (\R cm and \R cm);
   \draw[postaction={decorate}] (0.01,\R-0.21) arc (110:198:0.5*\R cm and \R cm);
   \draw[postaction={decorate}] (0.01,\R-0.21) arc (110:195:0.95*\R cm and \R cm);
      \draw[postaction={decorate}] (0.01,\R-0.21) arc (110:199:0.1*\R cm and \R cm);
 \node[anchor=north] at (-0.1,\R-0.6) {$^{t \rightarrow+\infty}$};
   \fill[fill=black] (0,\R-0.2) circle (1.2pt);
   \node[below=5pt] at (-1.2,-1.5){$|\phi_{i}^L\rangle
   $};
    \node[left=-2pt] at (-1.5,0.5){$|\phi_{i}^{L'}\rangle
    $};
 \node[below ] at (3.2,0) {(a)}; 
   \node[right] at (0.1,-1.25){$L$};
      \node[right] at (0.1,0) {$L'$};
       \clip (-\R+1.2,-\R) rectangle (0cm,1cm);
          \draw[very thick]  (-\R,1.45) arc (179.5:266.5:\R cm and 0.383*\R cm);
 \draw[very thick]  (-\R,0) arc (180:267:\R cm and 0.333*\R cm);
\end{tikzpicture}
\hspace{0.6cm}
\begin{tikzpicture}[scale=0.5, every node/.style={scale=0.65},rotate=270,decoration={
markings,
mark=between positions 0.1 and 0.9 step 8mm with {\arrowreversed{stealth}}} ]
\def\R{3} 
   \draw (0,0) circle  (\R cm and \R cm);
 \shade[ball color=blue!5!white,opacity=0.4] (-\R,0)   (0,0) circle  (\R cm and \R cm);
   \draw[postaction={decorate}] (0.01,\R-0.21) arc (110:198:0.5*\R cm and \R cm);
   \draw[postaction={decorate}] (0.01,\R-0.21) arc (110:195:0.95*\R cm and \R cm);
      \draw[postaction={decorate}] (0.01,\R-0.21) arc (110:199:0.1*\R cm and \R cm);
 \node[anchor=north] at (-0.1,\R-0.6) {$^{t \rightarrow+\infty}$};
   \fill[fill=black] (0,\R-0.2) circle (1.2pt);
 \node[below ] at (3.2,0) {(b)};    \node[left=5pt] at (-1.2,-0.8){$L$};
      \node[left=5pt,color=red] at (1,-0.8){$L^c$};
     \draw[dashed,thick, color=red] (-\R,0) arc (180:360:\R cm and 0.333*\R cm);
       \draw[dashed,thick, opacity=0.3,color=red] (-\R,0) arc (180:0:\R cm and 0.27*\R cm);   
       \clip (-\R+1.2,-\R) rectangle (0cm,1cm);
 \draw[very thick]  (-\R,0) arc (180:267:\R cm and 0.333*\R cm);
\end{tikzpicture}
\hspace{0.3cm}
\begin{tikzpicture}[scale=0.5, every node/.style={scale=0.65},rotate=270,decoration={
markings,
mark=between positions 0.1 and 0.9 step 8mm with {\arrowreversed{stealth}}} ]
\def\R{3} 
   \draw (-\R,0) arc  (180:360:\R cm and \R cm);
   \draw[dashed, very thick,color=red] (0,0) ellipse (\R cm and 0.333*\R cm);
     \fill[fill=gray!50] (0,0) ellipse (\R cm and 0.333*\R cm);
   \shade[ball color=blue!5!white,opacity=0.4] (-\R,0) arc  (180:360:\R cm and \R cm) -- (\R,0) arc (0:-180:\R cm and 0.333*\R cm);
 \node[left] at (-1,-\R+0.3) {$^{t\rightarrow-\infty}$};
 \node[below ] at (3.2,-1) {(c)}; 
   \node[right=5pt] at (-0.8,-2.5){$\langle\phi_{j}^{L}|$};
     \node[right=-4pt,color=red] at (1,-2.2){$\langle\phi^{L^c}|$};
\clip (-\R+1.2,-\R) rectangle (0cm,1cm);
 \draw[very thick]  (-\R,0) arc (180:267:\R cm and 0.333*\R cm);
\end{tikzpicture}
\hspace{-0.5cm}
\begin{tikzpicture}[scale=0.5, every node/.style={scale=0.65},rotate=270,decoration={
markings,
mark=between positions 0.1 and 0.9 step 8mm with {\arrowreversed{stealth}}} ]
\def\R{3} 
   \draw (\R,0) arc  (0:180:\R cm and \R cm);
   \draw[dashed,thick] (-\R,0) arc (180:360:\R cm and 0.333*\R cm);
   \shade[ball color=blue!5!white,opacity=0.4] (\R,0) arc  (0:180:\R cm and \R cm) -- (-\R,0) arc (180:360:\R cm and 0.333*\R cm);
    \draw[dashed,thick, opacity=0.3] (-\R,0) arc (180:0:\R cm and 0.27*\R cm);   
     \draw[postaction={decorate}] (0.01,\R-0.21) arc (110:198:0.5*\R cm and \R cm);
   \draw[postaction={decorate}] (0.01,\R-0.21) arc (110:195:0.95*\R cm and \R cm);
      \draw[postaction={decorate}] (0.01,\R-0.21) arc (110:199:0.1*\R cm and \R cm);
 \node[anchor=north] at (-0.1,\R-0.6) {$^{t \rightarrow+\infty}$};
   \fill[fill=black] (0,\R-0.2) circle (1.2pt);
  \node[below ] at (3.2,1) {(d)};   \node[left=0pt] at (-1.2,-0.8){$|\phi_{i}^{L}\rangle$};
       \node[left=1pt,color=red] at (1,-0.8){$|\phi^{L^c}\rangle$};
     \draw[dashed,thick, color=red] (-\R,0) arc (180:360:\R cm and 0.333*\R cm);
       \draw[dashed,thick, opacity=0.3,color=red] (-\R,0) arc (180:0:\R cm and 0.27*\R cm);   
    \clip (-\R+1.2,-\R) rectangle (0cm,1cm);
 \draw[very thick]  (-\R,0) arc (180:267:\R cm and 0.333*\R cm);
\end{tikzpicture}

\caption{ (a) Time development from section $L$ to $L'$ by the generator with $b^2-4ac>0$. (b) Section $L$ is complemented by the remainder of the space $L^c$, which  also develops into a fixed point as $t\rightarrow \infty$. The entire sphere can be divided into two hemispheres, (c) and (d), each boundary of which accommodates  states $\langle \phi_{j}^{L}|$ and $|\phi_{i}^{L}\rangle$ respectively. Gluing (c) and (d) at the red dashed line yields Eq. (\ref{eqn:1vacvac0}), or Fig. \ref{fig:Hmod} (a).}
\label{fig:LtoLdash}
\end{center}
\end{figure}
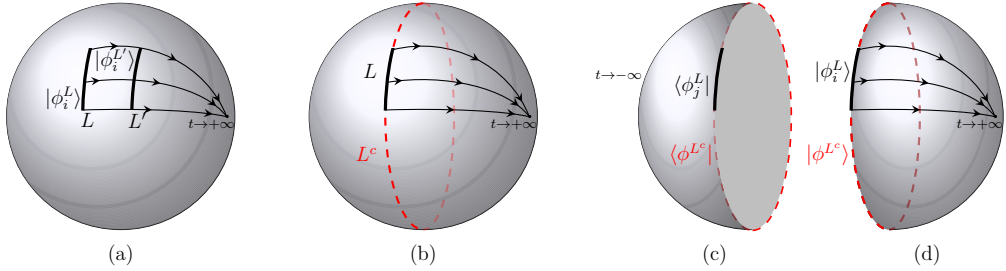

If the time-flow generated by $H$ has a fixed point as $t$ increases to infinity as illustrated in Fig. \ref{fig:LtoLdash} (a), 
any state can flow into the lowest energy state ({\it i.e.}, the vacuum):
\begin{equation}
|\phi_{i}^{L'}(t)\rangle =  e^{-tH}|\phi_{i}^L\rangle \rightarrow |0\rangle .
\end{equation}
The situation can be succinctly summarized in the following path integral:
\begin{equation}
\langle 0 | \phi^{L}_{i} \rangle=\int {\cal {D}} \phi^{L^c}(x^c) \int_{\phi(0,x)=\phi^{L}_{i}(x), \phi(0,x^c)=\phi^{L^c}(x^c)} {\cal D} \phi (t_{\geq 0},x)e^{-S}, \label{vacL0}
\end{equation}
where $L^c$ is the space compliment to $L$ (Fig. \ref{fig:LtoLdash} (b)). We attach either $L^c$ or superscript $^{c}$ to the functions, states, coordinates that are associated with $L^c$ (see Fig. \ref{fig:LtoLdash} (d)). In addition, $S$ is the appropriate conformal symmetric action. It should be noted that $t$ should take only positive values, hence the subscript ${\geq 0}$ is attached.
The Hermitian conjugation of Eq. (\ref{vacL0}) with a different state $| \phi^{L}_{j} \rangle$ on the $L$ can be written as
\begin{equation}
 \langle \phi^{L}_{j}| 0 \rangle=\int {\cal {D}} \phi^{L^c}(x^c) \int_{\phi(0,x)=\phi^{L}_{j}(x), \phi(0,x^c)=\phi^{L^c}(x^c)} {\cal D} \phi (t_{\leq 0},x)e^{-S}, \label{L1vac}
\end{equation}
where $t$ takes negative values (Fig. \ref{fig:LtoLdash} (c)).

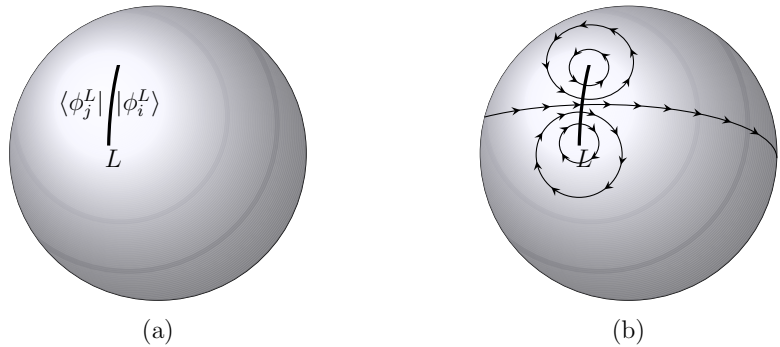
\begin{figure}[htbp]
\begin{center}
\begin{tikzpicture}[scale=0.65, every node/.style={scale=0.8},
rotate=270,decoration={
markings,
mark=between positions 0.1 and 0.9 step 4mm with {\arrowreversed{stealth}}} ]
\def\R{3} 
   \draw (0,0) circle  (\R cm and \R cm);
 \shade[ball color=blue!5!white,opacity=0.4] (-\R,0)   (0,0) circle  (\R cm and \R cm);
   \node[below=5pt] at (-1.7,-0.4){$|\phi_{i}^L\rangle $};
   \node[left=-2pt] at (-1,-1){$\langle \phi_{j}^{L}|$};
 \node[below ] at (3.2,0) {(a)}; 
   \node[right] at (0.1,-1.25){$L$};
       \clip (-\R+1.2,-\R) rectangle (0cm,1cm);
 \draw[very thick]  (-\R,0) arc (180:267:\R cm and 0.333*\R cm);
\end{tikzpicture}
\hspace{2cm}
\begin{tikzpicture}[scale=0.65, every node/.style={scale=0.8},
rotate=270,decoration={
markings,
mark=between positions 0.1 and 0.9 step 4mm with {\arrowreversed{stealth}}} ]
\def\R{3} 
   \draw (0,0) circle  (\R cm and \R cm);
 \shade[ball color=blue!5!white,opacity=0.4] (-\R,0)   (0,0) circle  (\R cm and \R cm);
   \draw[postaction={decorate}] (-0.2,-1) circle [x radius=0.4cm, y radius=0.4cm];
     \draw[postaction={decorate}] (0.9,-1.0) arc (0:360: 0.87 cm and 0.87cm);
   \draw[postaction={decorate}] (-1.38,-0.8) arc (360:0: 0.37 cm and 0.4cm);
  \draw[postaction={decorate}] (-1.1,-0.78) arc (360:0: 0.76 cm and 0.88cm);
   \draw[postaction={decorate}] (0.1,\R) arc (90:220:0.365*\R cm and 1.2*\R cm);
 \node[below ] at (3.2,0) {(b)}; 
   \node[right] at (0.1,-1.25){$L$};
       \clip (-\R+1.2,-\R) rectangle (0cm,1cm);
 \draw[very thick]  (-\R,0) arc (180:267:\R cm and 0.333*\R cm);
\end{tikzpicture}
\caption{(a) Cut $L$ on Riemann sphere. Two different states $\langle \phi^{L}_{j}|$ and $ | \phi^{L}_{i} \rangle$ are assigned on the cut.  (b) Time development that begins at $L$ and returns to $L$. The corresponding generator is called the modular Hamiltonian $H_{\mathrm{mod}}$. }
\label{fig:Hmod}
\end{center}
\end{figure}

One can the glue Eqs.  (\ref{vacL0}) and (\ref{L1vac}) ((c) and (d) of Fig.\ref{fig:LtoLdash} ) to obtain the following expression:
\begin{equation}
\langle \phi^{L}_{j}| 0 \rangle\langle 0 | \phi^{L}_{i} \rangle= \int_{\phi(0^{-},x)=\phi^{L}_{j}(x), \phi(0^{+},x)=\phi^{L}_{i}(x)} {\cal D} \phi (t,x)e^{-S}. \label{eqn:1vacvac0}
\end{equation}
Another way to explicate Eq. (\ref{eqn:1vacvac0}) is to consider a cut with  length $L$ on each side of which $\langle \phi^{L}_{j}|$ and $ | \phi^{L}_{i} \rangle$ reside respectively. The path integration is performed on the entire Riemann sphere with the exception of cut $L$ (Fig.\ref{fig:Hmod}(a)). One can convince oneself of Eq. (\ref{eqn:1vacvac0}) by integrating over  cut $L$ with the condition $\phi^{L}_{j}(x)=\phi^{L}_{i}(x)$, since the integration (or the trace) yields the partition function $\langle 0 | 0 \rangle$, which is simply the path integral over the entire sphere.

Note that the left-hand side of Eq. (\ref{eqn:1vacvac0}) takes the form of a density matrix that corresponds to the vacuum.  It is implicit that this density matrix depends only on the sector related to $L$, since the sector originating from the complement space $L^c$ is already integrated  in the right-hand side of Eq. (\ref{eqn:1vacvac0}). Therefore, this matrix is the reduced density matrix of the vacuum:
\begin{equation}
\langle \phi^{L}_{i}| 0 \rangle\langle 0 | \phi^{L}_{j} \rangle \equiv Z\times\left(\rho\right)_{ij} = \left( \mathrm{tr}_{L^c} \left( | 0 \rangle\langle 0 | \right) \right)_{ij}.
\end{equation}
where $Z$ is the partition function of the entire system:
\begin{equation}
Z = \sum_{i} \langle \phi^{L}_{i}| 0 \rangle\langle 0 | \phi^{L}_{i} \rangle=\sum_{i}\left( \mathrm{tr}_{L^c} \left( | 0 \rangle\langle 0 | \right) \right)_{ii}.
\end{equation}
The reduced density matrix $\rho$ is normalized as
\begin{equation}
\sum_{i}\left(\rho\right)_{ii}=1.
\end{equation}

If the reduced density matrix can be written as the exponentiation of a Hermitian operator,
\begin{equation}
\rho = \frac{e^{-TH_{\mathrm{mod}}}}{\mathrm{tr}\left( e^{-TH_{\mathrm{mod}}}\right)}=\frac{e^{-TH_{\mathrm{mod}}}}{Z} ,
\end{equation}
the operator is called the modular Hamiltonian in the context of axiomatic quantum field theory \cite{Haag:1992hx,1102.0440}. In the context of statistical physics, this operator is also called the entanglement Hamiltonian \cite{Hislop:1981uh,Li:2008kda,cond-mat/0410416,Casini:2011kv,Blanco:2013joa,1305.3291,Cardy:2016fqc,Nishioka:2018khk}.
For the case at hand,  the modular Hamiltonian in question is simply
\begin{equation}
H_{\mathrm{mod}}=aL_{1}+bL_{0}+cL_{-1} + a\bar L_{1}+b \bar L_{0}+c\bar L_{-1} = \L_{0}+\bar\L_{0},
\end{equation}
with $b^{2}-4ac \leq 0$, which we have been examining in this study. 
This is because we require the time flow that begins on one side of $L$, where the state $ | \phi^{L}_{i} \rangle$ is located, and returns to the other side of $L$ where $\langle \phi^{L}_{j}|$ is assigned, after covering the entire sphere.
See Fig. \ref{fig:Hmod} (b) and compare it with Fig. \ref{fig:timeflow2}.

With these setups, it is  almost trivial  to derive the entanglement 
entropy for section $L$. Following the treatment in \cite{Holzhey:1994we}, it is useful to introduce the following generalization of the partition function:
\begin{equation}
Z(n)  \equiv \mathrm{tr}e^{-nTH_{\mathrm{mod}}}={Z^{n}} \mathrm{tr} \rho^n,
\end{equation}
which includes the expression $ \mathrm{tr} \rho^n$. It is a well-known trick that the derivative of $ \mathrm{tr} \rho^n$ yields the entropy for the system governed by the density matrix $\rho$:
\begin{equation}
-\left. \frac{d}{dn}\mathrm{tr} \rho^n\right|_{n=1}=-\mathrm{tr}\left(\rho\ln\rho\right)=S.
\end{equation}
Because the reduced density matrix is considered, we obtain the following expression for the entanglement entropy:
\begin{equation}
S=-\left. \frac{d}{dn}\frac{Z(n)}{Z^{n}}\right|_{n=1}=\left. \left(1-n\frac{d}{dn}\right)\ln Z(n) \right|_{n=1}. \label{eqn:SnZn}
\end{equation}

The partition function $Z$ can be calculated through the path integral over the sphere; however, the existence of the cut-off alters the integral to that on the torus with the moduli parameter $\tau$ (\ref{eqn:modulidef}), as demonstrated in the previous section. By introducing
\begin{equation}
q=e^{2\pi i \tau},
\end{equation}
the partition function can be  expressed in the following familiar form:
\begin{equation}
Z_{\tau}=\mathrm{tr} q^{\L_{0}}\bar q^{\bar\L_{0}},
\end{equation}
where the subscript explicitly denotes the moduli parameter.
As $\L_{n}$ represents the Virasoro charge on the torus, there should be a corresponding Virasoro algebra  $L_{n}$ on the sphere, whose energy-momentum tensor differs  as
\begin{equation}
\L_{0}=L_{0}-\frac{\cc}{24} ,
\end{equation}
due to the Schwarzian derivative.
Therefore, the partition function can also be expressed as follows:
\begin{equation}
Z_{\tau}=q^{(-\frac{\ccs}{24})}\bar q^{(-\frac{\ccbs}{24})}\mathrm{tr} q^{L_{0}}\bar q^{\bar L_{0}}.
\end{equation}
To obtain $Z(n)$ or $Z_{\tau} (n)$, one can simply replace $q$ with $q^{n}$ in the above expression. Noting that the $n$ dependence only enters in the following combination,
\begin{equation}
\ln q^{n}=n\ln q\ ,\  \ln\bar q^{n}=n\ln \bar q ,
\end{equation}
one can replace the $n$ derivative in Eq. (\ref{eqn:SnZn}) with the derivative by $\ln q$ and $\ln \bar q$:
\begin{equation}
S=\left(1-\ln q \frac{\partial}{\partial \ln q}-\ln \bar q \frac{\partial}{\partial \ln \bar q}\right)\ln Z_{\tau}(\ln q, \ln \bar q). 
\end{equation}
One can further exploit the modular invariance of the partition function on the torus by the following modular transformation:
\begin{equation}
\tau \rightarrow -\frac{1}{\tau}.
\end{equation}
The expression for the entanglement entropy becomes
\begin{equation}
S=\left(1+\ln q \frac{\partial}{\partial \ln q}+\ln \bar q \frac{\partial}{\partial \ln \bar q}\right)\ln Z_{-\frac{1}{\tau}}(\ln q, \ln \bar q) , 
\end{equation}
and 
\begin{equation}
\ln q = 2\pi i \left(-\frac{1}{\tau}\right)=-2 \ln \frac{L}{\varepsilon}.
\end{equation}
Then, as argued  in \cite{Holzhey:1994we},  the contribution from term $\mathrm{tr} q^{L_{0}}$ to the partition function is exponentially suppressed provided that $L_{0}$ is a positive definite operator. Because the relevant contribution originates only from the part $q^{(-\frac{\ccs}{24})}\bar q^{(-\frac{\ccbs}{24})}$, we arrive at
\begin{equation}
S=\frac{\cc+\ccb}{6} \ln \frac{L}{\varepsilon},
\end{equation}
which is in accordance with the known results.

It would be instructive to recast the above analysis  in terms of the operator algebra language, such as Tomita-Takesaki theory \cite{Takesaki:1970aki}, the Reeh-Schlieder theorem \cite{Reeh:1961ujh, Redhead:1994pfg, Redhead:1995tf, Clifton:1997rz,1803.04993}, and the Bisognano-Wichmann theorem \cite{Bisognano:1975ih,Bisognano:1976za} in particular. We leave this fascinating subject for future study.



\section{Discussion}
Thus, we have established the relation between  our formalism and  entanglement entropy. In our treatment, the cut-off that is required in the expression of entanglement entropy is naturally and geometrically introduced. The boundary condition at the cut-off is also determined from the consistency of the Virasoro algebra. The entanglement entropy and the cut-off boundary condition have also been  discussed in the literature \cite{Lauchli:2013jga,Ohmori:2014eia,Herzog:2016bhv}.

Another interesting association with our formalism can be seen by considering the infinitesimal limit of section $L$:
\begin{equation}
L=\frac{\sqrt{|b^{2}-4ac|}}{a} \rightarrow 0.
\end{equation}
This limit can be achieved by, for example, taking $b\rightarrow 1$ while retaining $a=c=-\frac12$, which in the limit yields
\begin{equation}
g(z)=-\frac12 z^{2}+z-\frac12=-\frac12(z-1)^{2} ,
\end{equation}
and the following time development operator:
\begin{equation}
L_{0}-\frac{L_{1}+L_{-1}}{2}+\bar L_{0}-\frac{\bar L_{1}+\bar L_{-1}}{2}.
\end{equation}
The above time development operator is simply the SSD Hamiltonian \cite{Katsura:2011ss,Ishibashi:2015jba,Ishibashi:2016bey}.

To further investigate this limit, it is rather convenient to retain the original notion of $\kappa $ in Eq. (\ref{eqn:kappaLn}) instead of rescaling by $\xi$. It is then apparent from Eq. (\ref{eqn:kappaLn}) that  $\kappa$ takes continuous values in the $L\rightarrow 0$ limit. The factor between $\kappa$ and  integer $\frac{\pi aL}{\ln\left({L}/{\varepsilon}\right)}$ also appears as the (inverse) factor in the expression of the central charge (\ref{eqn:cikmk}), yielding the delta function in the $L\rightarrow 0$ limit. Therefore, simply by taking the  $L\rightarrow 0$ limit, we obtain the continuous Virasoro algebra, which was found in \cite{Ishibashi:2015jba,Ishibashi:2016bey}:
\begin{equation}
[\L_{\kappa} , \L_{\kappa'} ]=(\kappa-\kappa')  \L_{\kappa+\kappa'}+\frac{\cc}{12}\kappa^{3}\delta(\kappa+\kappa') .    \label{VirxiSSD}
\end{equation}
Here, we also take the step of  shifting $\L_{0}$ as Eq. (\ref{eqn:L0shift}). 

In this limit, 
the new structure of the continuous Virasoro algebra emerges. The details of this limiting procedure will be addressed in future publications. It would be interesting to explore the implication of the SSD Hamiltonian and  continuous Virasoro structure by taking the $L\rightarrow 0$ limit in the study of entanglement entropy.

\begin{figure}[htbp]
\begin{center}
\begin{tikzpicture}[scale=0.65, every node/.style={scale=0.75},rotate=270,decoration={
markings,
mark=between positions 0.1 and 0.9 step 4mm with {\arrowreversed{stealth}}}]
\def\R{3} 
   \draw (0,0) circle  (\R cm and \R cm);
 \shade[ball color=white!90!black,opacity=0.4, shading angle=220] (\R,\R)   (0,0) circle  (\R cm and \R cm);
   \draw[postaction={decorate}] (-0.2,-1) circle [x radius=0.4cm, y radius=0.4cm];
    \draw[densely dotted,thick, pattern=north west lines
    ] (-0.2,-1) circle [x radius=0.32cm, y radius=0.32cm];
      \draw[densely dotted,thick, pattern=north east lines] (-1.75,-0.8) circle [x radius=0.28cm, y radius=0.3cm];
     \draw[postaction={decorate}] (0.9,-1.0) arc (0:360: 0.87 cm and 0.87cm);
   \draw[postaction={decorate}] (-1.38,-0.8) arc (360:0: 0.37 cm and 0.4cm);
  \draw[postaction={decorate}] (-1.1,-0.78) arc (360:0: 0.76 cm and 0.88cm);
   \draw[postaction={decorate}] (0.1,\R) arc (90:220:0.365*\R cm and 1.2*\R cm);

       \clip (-\R+1.2,-\R) rectangle (0cm,1cm);

\end{tikzpicture}

\caption{The hatched disks are removed from the sphere, and their boundaries (dotted circles) are identified with each other, changing the topology from a sphere to a torus. Note, in particular, that the shape of each boundary of the disks is congruent with one of the time flows (arrowed lines).}
\label{Fig:Torus-construction}
\end{center}
\end{figure}
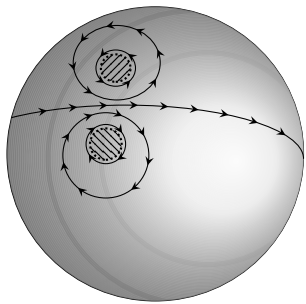

The formalism developed in this study has a wider and  intriguing application. In the course of the above analysis,  the existence of the cut-off, which itself is introduced  to preserve the consistency of the Virasoro algebra, leads us to the Virasoro algebra on a torus. Therefore, the same analysis with an arbitrary large ``cut-off '' should yield the Virasoro algebra on a torus with an arbitrary moduli parameter. This also opens up the possibility of constructing the Virasoro algebra on a general two-dimensional surface with the higher genus than the sphere and torus by contriving an appropriate time-flow and applying the gluing procedure, as illustrated in Fig. \ref{Fig:Torus-construction}. It would be also interesting to apply the present formalism to the calculation of the entanglement entropy for multiple sections \cite{Calabrese:2004eu, Furukawa:2008uk, Calabrese:2009ez,1303.6955,1608.01283,1805.10651}. These possibilities will be pursued in future research.

In summary, we have shown that the time-flow associated with the class of $L_{1}+L_{-1}$ operator leads to the Virasoro algebra on a torus. This fact was utilized to re-derive the entanglement entropy in a rather straightforward manner without resorting to a contrived transformation.

\vspace{2cm}
\noindent{\bf Acknowledgment:} The author would like to thank N. Ishibashi for the collaboration at the early stage of the research. His gratitude is also extended to E. Itou, H. Katsura, H. Kawai, Y. Kubota, M. Nozaki, K. Okunishi, S. Ryu, T. Takahashi, 
X. Wen, G. Wong, 
 and the participants of the iTHEMS workshop ``Workshop on Sine square deformation and related topics,'' for fruitful discussions,  comments and suggestions, which greatly contributed to the present work. The author would also like to thank J. Polchinski for his memoirs \cite{Polchinski:2017vik}, reading which brought the author fond memories, encouragement, inspiration, and some consolation during the course of the research presented here.


\end{document}